\documentclass[prb,preprint,showpacs,preprintnumbers,amsmath,amssymb]{revtex4}

\usepackage{longtable}
\usepackage{graphicx}
\usepackage{dcolumn}
\usepackage{bm}
\graphicspath{{.},{./Ps/},{../Ps/},{../../Ps/}}                                            

\begin{document}

\title{Morphology changes in the evolution of
 liquid two-layer films}

\author{Andrey Pototsky}
\affiliation{Lehrstuhl f{\"u}r Theoretische Physik II,
 Brandenburgische Technische Universit\"at Cottbus,
 Erich-Weinert-Stra{\ss}e 1, D-03046 Cottbus, Germany}

\author{Michael Bestehorn}
\affiliation{Lehrstuhl f{\"u}r Theoretische Physik II,
 Brandenburgische Technische Universit\"at Cottbus,
 Erich-Weinert-Stra{\ss}e 1, D-03046 Cottbus, Germany}

\author{Domnic Merkt}
\affiliation{Lehrstuhl f{\"u}r Theoretische Physik II,
 Brandenburgische Technische Universit\"at Cottbus,
 Erich-Weinert-Stra{\ss}e 1, D-03046 Cottbus, Germany}

\author{Uwe Thiele}
\affiliation{Max-Planck-Institut f\"ur Physik komplexer Systeme,
 N\"othnitzer Stra{\ss}e 38, D-01187 Dresden, Germany}

\begin{abstract}
We consider a thin film consisting of two layers of immiscible liquids on a 
solid horizontal (heated) substrate. 
Both, the free liquid-liquid and the liquid-gas interface of such a
bilayer liquid film may be unstable due to effective molecular interactions 
relevant for ultrathin layers below 100\,nm thickness,  
or due to temperature-gradient caused Marangoni flows in the heated case. 
Using a long wave approximation we derive coupled evolution equations 
for the interface profiles for the general non-isothermal situation
allowing for slip at the substrate. Linear and nonlinear analyses 
of the short- and long-time film evolution are
performed for isothermal ultrathin layers taking into account
destabilizing long-range and stabilizing short-range molecular interactions.
It is shown that the initial instability can be of
a varicose, zigzag or mixed type. However, in the nonlinear stage of the evolution
the mode type and therefore the pattern morphology can change 
via switching between two different branches of stationary solutions 
or via coarsening along a single branch.
\end{abstract}

\maketitle

\section{Introduction.}
\label{intro}
Instabilities of thin liquid films between a solid substrate and a gas atmosphere
have attracted much scientific interest. The main focus lies thereby on front
instabilities of moving contact lines \cite{CHTC90,BrGe93,BMFC98} or on instabilities of the 
free liquid-gas interface of a flat film \cite{Reit92,VanH95,TMP98}. 
A recent review can be found in Ref.\,\onlinecite{ODB97}.
To analyze such instabilities a long-wave or 
lubrication approximation \cite{Reyn1886,Somm1904,ODB97} is often used as a very powerful tool
especially for low Reynolds number film flows.
At present the basic behavior of one-layer films 
in the physically different thickness ranges is well understood. 
Several instability mechanisms exist that by means of different driving forces
may destabilize an initially flat film. They are described, analysed and modelled in a large
number of experimental \cite{Reit92,ShRe96,Reit93a,JSSH98,VanH95,VanH97,SHJ01}
and theoretical 
\cite{RuJa73,WiDa82,deGe85,YiHi89,OrRo92,Shar93,Shar93b,Mitl93,ODB97,
ShKh98,TVN01,ScSt02,TNPV02,Thie03,BGW01,BPT03,ThKn04,PBT04}
works. For film thicknesses $d$ less than about $100$\,nm, effective 
molecular interactions between the film surface and the substrate 
dominate all the other forces, like thermo- and soluto-capillarity or gravity,
and thus determine the film stability.
For heated films of thicknesses above $100$\,nm, eventually thermocapillary
forces become the most important influence leading to an instability 
caused by large-scale Marangoni convection \cite{VanH95,BPT03}.
It is dominant up to an upper limit of the film thickness determined
by the competition between large-scale and small-scale convection modes
\cite{GNP94}. 
For even thicker films with thicknesses above $100\,\mu$m also the gravity force becomes
important. Depending on its direction it may stabilize the large-scale Marangoni instability or 
destabilize the film further  (Rayleigh-Taylor instability) \cite{YiHi89,OrRo92}.
The lubrication approximation is valid up to a limiting film thickness
obtained by the requirement that the wave length of the 
dominant instability mode $\lambda_m$ is much larger than the film thickness
$d$, i.e.\ $\lambda_m\gg d$. For the Rayleigh-Taylor instability $\lambda_m$
depends on the interfacial tension, 
the density of the liquid and the gravitational acceleration but not 
on the film thickness \cite{ODB97}. It is of the order of 
$10^{3}$ to $10^{4}\,\mu$m implying an upper limit for the film thickness 
of $10^2$ to $10^{3}\,\mu$m.

The stability and evolution of liquid films
of thicknesses below $100$\,nm is determined by effectiv 
molecular interactions between substrate and film arising, for instance, from
Van der Waals, electrostatic or entropic interactions \cite{Isra92,Hunt92}.
Such films are lineary unstable if the energy of the intermolecular 
interaction is a convex function of the film thickness.
For film thicknesses above $10$\,nm the long-range 
Van der Waals forces dominate. 
They can be of different nature depending on the molecular properties of the
involved media. One distinguishes interactions between two randomly orienting dipoles 
(orientation interaction), between a randomly 
orienting dipole and an induced dipole (induction interaction), and between 
a fluctuating dipole and an induced dipole (dispersion interaction).
Between two parallel interfaces at a distance $d$, all
these forces decay as $A/d^3$ where $A$ is the Hamaker constant \cite{Isra92}.
An unstable situation corresponds to a positive Hamaker constant. Note,
however that different schools use different sign conventions. 
The dominant wave length of the instability $\lambda_m$ increases 
monotonically with $d$ as $\lambda \sim d^2$ (see Ref.\,\onlinecite{ODB97}).
The stability of a film may change dramatically for a substrate 
coated with a layer of different dielectric
properties as, for instance, a silicon substrate (Si) coated with an silicon 
oxide layer (SiO) \cite{SHJ01c}. There, for an oxide layer of about
2\,nm only ultrathin polystyrene (PS) films below 4\,nm thickness are linearly unstable. 
Increasing the film thickness in the linearly unstable range, 
the wave length $\lambda$ increases rapidly and diverges at the critical
thickness $d_c$. For $d>d_c$ the film is linearly stable, but may rupture due
to finite disturbances.

Imagine one replaces the (solid) coating layer by a liquid layer transforming
the system in a two-layer liquid film. Some of the results obtained for a
solid coating can be directly transferred to the new situation.
The stability of the (now) upper layer still depends on the (now liquid) coating layer.
However, additionally the liquid coating layer itself may be unstable making a
re-evaluation of the stability necessary. 
This thought experiment leads quite naturally to the extension of the well 
studied one-layer systems to two-layer systems that are the subject of the
present work.
In general, there exist two possible two-layer geometries. On the one hand
the two liquid layers can be situated
between two solid plates leaving only the interface between the two liquids
free to move. In consequence such a system can be described by a single
evolution equation \cite{Lin01,Lin02,MPBT05}.
On the other hand the two layers can be situated between a solid substrate and
a gas atmosphere. Then both, 
the liquid-liquid and the liquid-gas interface are free to move and their evolution
has to be described by coupled evolution equations. Models were derived, for
instance, assuming a lower liquid layer that is much thicker than the upper
layer \cite{BMR93}, and for two-layer systems with surfactants (and non-newtonian behaviour)
\cite{ZMC03,CrMa00,MCW02} or including 
evaporation \cite{Dano98,Dano98b,Paun98}. A two-layer system under the solely
influence of molecular interactions is studied in
Ref.\,\onlinecite{Band01}. In Ref.\,\onlinecite{PBMT04} a similar system is
studied, however, the evolution equations are given in terms of variations of an energy 
functional. 

The experimental interest in two-layer liquid films is up to now mainly focused 
on the dewetting of a liquid layer from a very thick layer, i.e.\ a liquid bulk substrate 
\cite{Wunn03,MSS04,FCW95,LPHK96,PWHC97}.
In contrast Ref.\,\onlinecite{RMSH00} studies
the dewetting of a polystyrene (PS) layer of $15$ to $68$\,nm thickness from a 
$46$\,nm thick polyamide (PA) layer. The substrate is a silicon (Si) wafer
covered with a layer of native oxide.
At high temperature ($195^o$C) and small thicknesses ($15...35$\,nm)
the PS layer is unstable and dewets exhibiting typical spinodal patterns.
At low temperature ($115^o$C) the PA layer is solid resulting in a stable
PS layer, independent of its thickness. 

Ref.\,\onlinecite{MBB94} studies relatively thick layers ($100$\,nm to 
$1\,\mu$m thickness) of poly(dimethylsiloxane) (PDMS) layers on 
a liquid substrate of fluorinated PDMS. They show that the PDMS films 
are metastable and may dewet by nucleation of holes. 
The velocity of the growth of holes depends on the viscosity and thickness of 
the substrate. In another system (PS layer on poly(methylmethacrylate) (PMMA) layer, 
both with thicknesses of about $100$\,nm) the dewetting velocity was found to
exhibit a minimum as a function of the viscosity of the lower layer \cite{LPHK96}.
Furthermore, for a polycarbonate (PC) layer on a 
poly(styrene-$co$-acrylonitrile) (SAN) layer
Ref.\,\onlinecite{PWHC97} reports a non-trivial dependence of the 
dewetting velocity on both layer thicknesses. 

Recently we presented coupled evolution equations for 
two-layer liquid films taking into account long-range Van der Waals
interactions only \cite{PBMT04}. This allows to study pathways towards
rupture but can not describe the long-time evolution of such films as, for
instance, necessary for the description of the above mentioned experiments. 
In the present paper we present the derivation of the system of evolution
equations for a general interaction energy and a non-isothermal situation
and then study the short- and long-time evolution of
two-layer liquid films incorporating long-range and short-range interactions.
Thereby the main focus lies on ultrathin layers
with respective thicknesses below $100$\,nm for which the effective 
molecular interactions between the four media are the dominant influence.

The paper is structured as follows. The problem is formulated 
in Section~\ref{deriv} followed by the derivation of the relevant long wave
evolution equations for a general non-isothermal case allowing also for slip
at the substrate. Focusing on the isothermal case without slip these 
equations are analyzed and integrated numerically in the subsequent sections.
Section~\ref{lin_stab} presents the linear stability analysis for flat films starting from 
the general case in Section~\ref{gen_stab}, focusing on long-range van der
Waals interactions in Section~\ref{lr_int}, discussing the possible
mode types in Section~\ref{inst_modes} and introducing important limiting
cases in Section~\ref{lim_cases}. 
In Section~\ref{nuss} we investigate the non-linear behaviour 
discussing in Section~\ref{nu_sol} non-uniform stationary solutions
as extrema of the Lyapunov functional, presenting
in Section~\ref{mt_trans} two different pathways for mode-changes in the
course of the time evolution, and discussing long-time stationary
solutions for different 
experimentally relevant systems in Section~\ref{lp_sol}.
Concluding remarks follow in Section~\ref{conc}. Expressions for the 
surface tension gradients in terms of gradients of the layer thicknesses 
are discussed in an Appendix.
\section{Derivation of coupled large-amplitude evolution equations}
\subsection{General case}
\label{deriv}
First, we derive coupled evolution equations for the profiles 
of the liquid-liquid interface $h_1(x,y)$ and the liquid-gas interface 
$h_2(x,y)$. We assume that the layers are thin enough that
convective terms can be neglected. Considering a two-dimensional geometry as 
sketched in Fig.\,\ref{fig1} the respective 
Stokes equations for the two layers are
\begin{equation}
\nabla (p_i + \phi_i) = \mu_i \Delta \vec{v}_i,
\label{stokes}
\end{equation}
where $i=1,2$ denotes the respective layer. For each layer
$\vec{v_i} = (u_i,w_i)$ is the velocity field,
$p_i$ the pressure, $\phi_i$ the potential
of the bulk forces and $\mu_i$ the viscosity. The constant mean film
thicknesses are denoted by $d_1 = (\int_0^L h_1 dx)/L$ 
and $d_2 = (\int_0^L h_2 dx)/L$ where $L$ is the lateral system size.
A lubrication approximation \cite{ODB97} is applied assuming the ratio of
vertical and horizontal length scales to be small.
As smallness parameter we use the ratio $\epsilon = d_1 / \lambda$ where
$\lambda$ is the characteristic lateral length scale of the film instability.
In zeroth order in $\epsilon$ the Stokes equations 
(\ref{stokes}) simplify to
\begin{eqnarray}
\mu_2 \partial_z^2 u_2 = \partial_x \bar{p}_2 \label{NSa} \\
\partial_z \bar{p}_2 = 0 \label{NSb} \\
\mu_1 \partial_z^2 u_1 = \partial_x \bar{p}_1 \label{NSc} \\
\partial_z \bar{p}_1 = 0, 
\label{NSd}
\end{eqnarray}
where the $\bar{p}_i$ stand for $p_i + \phi_i$.
At the substrate $(z=0)$ we use a Navier slip and a no-penetration condition,
i.e.\
\begin{eqnarray}
u_1 = \beta \partial_z u_1~~\text{and}~~w_1 = 0,
\label{SLIP}
\end{eqnarray}
respectively. The slip length is denoted by $\beta$.
At the liquid-liquid interface $(z = h_1)$ we use the continuity of the 
velocity field, the kinematic condition and the continuity of the 
tangential component of the liquid stress tensor
\begin{eqnarray}
u_1 = u_2,~w_1 = w_2,  
\label{BCVa} \\
w_1 = \partial_t h_1 +u_1 \partial_x h_1,  
\label{BCVb} \\
\end{eqnarray}
and
\begin{eqnarray}
\mu_1 \partial_z u_1 - \mu_2 \partial_z u_2 = \partial_x \sigma_{12},  
\label{BCVc}
\end{eqnarray}
respectively. The normal stress condition is discussed below.
At the liquid-gas interface $(z = h_2)$ only the kinematic condition 
and the continuity of the tangential component of the liquid stress tensor
apply, i.e.\
\begin{eqnarray}
w_2 = \partial_t h_2 +u_2 \partial_x h_2, 
\label{BCV1a} \\
\mu_2 \partial_z u_2 = \partial_x \sigma_{2}. 
\label{BCV1b}
\end{eqnarray}
The $\sigma_{12}$ and $\sigma_2$ stand for the interfacial tensions of
the liquid-liquid and of the liquid-gas interface, respectively.
The boundary conditions for the normal component of the stress tensor
are written incorporating the disjoining pressures at the liquid-liquid
$\Pi_1(h_1,h_2)$ and at the liquid-gas $\Pi_2(h_1,h_2)$ interface,
respectively. They represent effective molecular
interactions between the interfaces that result, for instance, from Van der
Waals interactions \cite{Isra92}. They are discussed in detail below.
For the liquid-gas interface $(z=h_2)$ we obtain
\begin{eqnarray}
p_2(h_2) - p_0 = - \sigma_2 \partial_x^2 h_2 + \Pi_2(h_1,h_2)
\nonumber 
\end{eqnarray}
and for the liquid-liquid interface $(z=h_1)$
\begin{eqnarray}
p_1(h_1) - p_2(h_1) = - \sigma_{12} \partial_x^2 h_1 + \Pi_1(h_1,h_2),
\label{BCP}
\end{eqnarray}
where $p_0$ is the constant pressure in the gas atmosphere.
Eqs.\,(\ref{BCP}) can be written in terms of variations of an energy
functional
$F[h_1,h_2]$
\begin{eqnarray}
p_1(h_1)-p_2(h_1) = \frac{\delta F}{\delta h_1} \nonumber \\
p_2(h_2)-p_0 =  \frac{\delta F}{\delta h_2},
\label{BCP_ENRG}
\end{eqnarray}
with
\begin{equation}
F = \int \left(\sigma_{1} \frac{(\partial_x h_1)^2}{2} + 
\sigma_2 \frac{(\partial_x h_2)^2}{2} +f(h_1,h_2)
 \right)dx,
\label{LYAP}
\end{equation}
and $f(h_1,h_2)$ being the free energy of the flat films per unit area.

Equations (\ref{NSa}) and (\ref{NSc}) are integrated three times 
with respect to $z$ to obtain the stream functions $\Psi_i$, defined by
$(w_i = - \partial_x \Psi_i,~u_i = \partial_z \Psi_i)$. 
The six $x$-dependent integration constants
are determined using the boundary conditions  
(\ref{SLIP}), (\ref{BCVa}), (\ref{BCVc}), and (\ref{BCV1b}).
Thus the velocity fields in the two layers are given by
\begin{eqnarray}
u_1 &=& \frac{1}{\mu_1} \left( \partial_x \bar{p}_1\right)  \frac{z^2}{2}+
\frac{1}{\mu_1}(z+\beta) K_1 \nonumber \\
u_2 &=& \frac{1}{\mu_2} \left( \partial_x \bar{p}_2\right)  \frac{z^2}{2}+
\frac{1}{\mu_2} K_2 (z-h_1)-
\frac{\partial_x \bar{p}_2}{\mu_2} \frac{h_1^2}{2} + u_1 (h_1),
\label{velocities}
\end{eqnarray}
with $K_1 =  K_2 + \partial_x \sigma_{12} +
\left[(\partial_x \bar{p}_2)- (\partial_x \bar{p}_1)\right] h_1$ and 
$K_2 =  \partial_x \sigma_{2} - (\partial_x \bar{p}_2)h_2$.

The stream functions $\Psi_i$ are related to the flow in the lower layer
$\Gamma_1 = \int_{0}^{h_1} u_1\,dz$ 
and to the one in the upper layer 
$\Gamma_2 = \int_{h_1}^{h_2} u_2\,dz$ by
$\Gamma_1 = \Psi_1(h_1),~\Gamma_1 + \Gamma_2 = \Psi_2(h_2)$.
Using the $\Psi_i$ we rewrite Eqs.\,(\ref{BCVb}) and 
(\ref{BCV1a}) to obtain the evolution equations for the
two interface profiles
\begin{eqnarray}
\partial_t h_1 +\partial_x \left[ \Psi_1 (h_1) \right] &=& 0,
\label{EVOLH1} \\
\partial_t h_2 +\partial_x \left[ \Psi_2 (h_2) \right] &=& 0.
\label{EVOLH2}
\end{eqnarray}
Written in terms of the energy functional they read
\begin{eqnarray}
\partial_t h_1 &=& \partial_x \left[ Q_{11}\partial_x 
\frac{\delta F}{\delta h_1} 
+ Q_{12}\partial_x \frac{\delta F}{\delta h_2} - 
D_{11} \partial_x \sigma_{12} - D_{12} \partial_x \sigma_{2} \right] 
\nonumber \\ 
\partial_t h_2 &=& \partial_x \left[ Q_{21}\partial_x 
\frac{\delta F}{\delta h_1} 
+ Q_{22}\partial_x \frac{\delta F}{\delta h_2} -
 D_{21} \partial_x \sigma_{12}  -  D_{22} \partial_x \sigma_{2} \right],
\label{EVOL_PHYS}
\end{eqnarray}
with the mobility matrices of the pressure terms
\begin{eqnarray}
{\bm Q} = \frac{1}{\mu_1}
\left(
\begin{array}{ccc}
\frac{h_1^3}{3}+ \beta h_1^2 &~& \frac{h_1^2}{2}(h_2 -\frac{h_1}{3}) + 
\beta h_1 h_2\\
\frac{h_1^2}{2}(h_2 -\frac{h_1}{3})  + \beta h_1 h_2 &~& 
\frac{(h_2-h_1)^3}{3}(\frac{\mu_1}{\mu_2}-1) + \frac{h_2^3}{3} +\beta h_2^2
\end{array}
\right)
\label{MOB}
\end{eqnarray}
and of the tangential stress terms
\begin{eqnarray}
{\bm D} = \frac{1}{\mu_1}
\left(
\begin{array}{ccc}
\frac{h_1^2}{2}+ \beta h_1 &~& \frac{h_1^2}{2} + \beta h_1 \\
h_1(h_2 -\frac{h_1}{2})  + \beta h_2 &~& \frac{\mu_1 (h_2-h_1)^2}{2 \mu_2} + 
h_1(h_2 -\frac{h_1}{2}) 
+\beta h_2
\end{array}
\right),
\label{TANG}
\end{eqnarray}
respectively.
Note, that the mobility matrix ${\bm Q}$ is symmetric and all 
mobilities $Q_{ik}$ and $D_{ik}$ are positive. Dropping the terms representing
the effective molecular interactions, Eqs.\,(\ref{EVOL_PHYS})
represent the fully nonlinear equivalent for the weakly nonlinear equations derived in
Ref.\,\onlinecite{NeSi90,NeSi97}.
Assuming that the interfacial tensions are influenced by thermocapillarity 
only, one can express the derivatives
$\partial_x \sigma_{12}$ and $\partial_x \sigma_{2}$ in terms of gradients
of local thicknesses $\partial_x h_i$. This is done in the Appendix.

For isothermal ultrathin liquid films one has
$(\partial_x \sigma_{12} = \partial_x \sigma_{2} =0)$. The situation is then
relaxational (or variational), i.e.\ Eqs.\,(\ref{EVOL_PHYS}) 
possess a Lyapunov functional, namely the energy functional $F$, which
decreases monotonously in time as shown next.
The total time derivative of the Lyapunov functional is 
$dF/dt = \int \left( \frac{\delta F}{\delta h_1} \partial_ t h_1 + 
\frac{\delta F}{\delta h_2} \partial_ t h_2 \right) \,dx $.
Expressing $\partial_t h_i$ by Eqs.\,(\ref{EVOL_PHYS}) and 
using partial integration with periodic boundary conditions, one obtains
\begin{equation}
\frac{dF}{dt} = - \int \sum_{i,k} Q_{ik} \left( \partial_x 
\frac{\delta F}{\delta h_i} \right) 
\left(\partial_x \frac{\delta F}{\delta h_k} \right)\,dx. 
\label{LYAP_dt}
\end{equation}
Because 
\begin{equation}
\det {\bm Q} = \frac{(h_2-h_1)^3 h_1^3}{9 \mu_1 \mu_2} +\frac{1}{12 \mu_1^2} 
h_1^4 (h_2-h_1)^2
+ \beta \left( \frac{h_1^3}{3 \mu_1^2}(h_2-h_1)^2 +
h_1^2 \frac{(h_2-h_1)^3}{3 \mu_1 \mu_2} \right) > 0
\label{DET}
\end{equation}
and $Q_{ii} > 0$, the quadratic form in Eq.\,(\ref{LYAP_dt}) is positive 
definite implying $dF/dt < 0$.
The existence of $F$ allows to identify the stationary solutions of 
Eqs.\,\ref{EVOL_PHYS} with the extrema of $F$. 
This will be used below in Section\,\ref{nu_sol}.

\subsection{The disjoining pressures $\Pi_i(h_1,h_2)$}
\label{driv_for}
 In many important cases, as for instance, for polymer films on apolar 
 substrates \cite{SHJ01,Reit92}, the interaction energy is mainly 
 determined by its long-range apolar dispersion part.
 However, if the model only takes into account a destabilizing long-range interaction
 the time evolution definitively leads to rupture of 
 the upper or lower layer \cite{PBMT04} making it impossible to study the
long-time coarsening behaviour. To be able to study the long-time evolution
one has to include  stabilizing short-range interactions \cite{Isra92,Shar93}
into the model. Although these are normally not included for films of thicknesses
above $10$\,nm because they do not change the stability of flat films, also
for such films they become important in the non-linear stage of evolution
when the local thicknesses become comparable to their interaction length.

The long-range part of the interaction energy for each pair of interfaces
(see Fig.\,\ref{fig1}) resulting from dispersive Van der Waals interactions
is given by $A_{ijkl}/2 \pi h^2$ (see Ref.\,\onlinecite{Isra92}), 
where $A_{ijkl}$ is a
(four-index) Hamaker constant which corresponds to the interaction between
the interfaces $i-j$ and $k-l$.
Each index in $A_{ijkl}$ can be one out of $g,1,2$ and $s$,
denoting gas, liquid $1$, liquid $2$ and substrate, respectively.
The four-index Hamaker constant is calculated using
   an equivalent of Eq.\,(11.13) of Ref.\,\onlinecite{Isra92} that is based on the
   assumption that the main absorption frequencies of all involved media are
   about $\nu_e = 3\times10^{15}$\,Hz and that the zero frequency contribution 
 is negligible. One uses 
 \begin{equation}
   A_{ijkl}\,\approx\,\frac{3h\nu_e}{8\sqrt{2}}\,
   \frac{(n_i^2-n_j^2)(n_l^2-n_k^2)}{(n_i^2+n_j^2)^{1/2}(n_l^2+n_k^2)^{1/2}
   [(n_i^2+n_j^2)^{1/2}+(n_l^2+n_k^2)^{1/2}]}, 
 \label{HAM}
 \end{equation}
where $n_i$ are the refractive indices of the media.
The three-index Hamaker constants are given by
\begin{equation}
A_{ijk} = A_{ijjk}.
\label{HAMb}
\end{equation}

The short-range forces which can be of an electrostatic or structural nature 
\cite{TDS88,deGe85} decay exponentially with $h$.
The electrostatic part results from the formation of diffuse electric double-layers
in the vicinity of interfaces involving polar liquids
\cite{DeLa41,VeOv48,Ohsh74,Ohsh74b}. 
For films with thicknesses in the range of the Debye length 
the diffuse double-layers at the two interfaces overlap
resulting in a repulsive or attractive force between the interfaces.
The corresponding interaction energy between the interfaces $s-1$ and $1-2$ 
is given by $S_1 \exp{\left[ (l_0-h_1)/l_1\right]}$ and between the interfaces
$1-2$ and $2-g$ by $S_2 \exp{\left[ (l_0-(h_2-h_1))/l_2\right]}$, where
$l_0 = 0.158$\,nm is the Born repulsion length, and $l_1,~l_2 \sim 1..10$\,nm
are the interaction lengths of the short-range interactions \cite{Shar93}. 
Further on we consider the two correlation lengths $l_1$ and $l_2$ to be equal 
and denote them by $l$.
$S_1>0$ and $S_2>0$ are the short-range components of 
the total spreading coefficients. They are related to the lower layer on the
substrate  below a bulk of the upper liquid and to the upper layer
on the lower film as substrate below the ambient gas, respectively.
We do not take into account short-range interactions between interfaces $s-1$
and $2-g$.

Collecting the long-range and the short-range forces the
disjoining pressures $\Pi_1(h_1,h_2)$ and $\Pi_2(h_1,h_2)$ are specified as
 \begin{eqnarray}
\Pi_1(h_1,h_2)&=&\frac{A_{21s}}{6 \pi h_1^3} - 
 \frac{A_{12g}}{6 \pi (h_2-h_1)^3}- 
 \frac{S_1}{l_1} \exp{\left[\frac{l_0 - h_1}{l_1}\right]}+
 \frac{S_2}{l_2} \exp{\left[\frac{l_0 - (h_2-h_1)}{l_2}\right]}
\nonumber \\
 \Pi_2(h_1,h_2)&=&\frac{A_{12g}}{6 \pi (h_2 -h_1)^3} 
 +\frac{A_{g21s}}{6 \pi h_2^3}- 
 \frac{S_2}{l_2} \exp{\left[\frac{l_0 - (h_2-h_1)}{l_2}\right]}
 \label{DSJPRESS}
 \end{eqnarray}
To non-dimensionalize Eqs.\,(\ref{EVOL_PHYS}) we scale
the thicknesses with $l$, the lateral coordinate $x$ with 
$\lambda = l (d_2-d_1) \sqrt{2 \pi \sigma_1 /|A_{12g}|}$,
and time $t$ with $\tau = (2 \pi)^2 \sigma_1 \mu_1 l (d_2-d_1)^4 / A_{12g}^2$.
Then the scaled energy functional
\begin{eqnarray}
F &=& \int\,\left[ \frac{1}{2}(\partial_x h_1)^2 +\frac{\sigma}{2} (\partial_x h_2)^2
 -\frac{\bar{A}_{12g}}{6 (h_2-h_1)^2}-
\frac{\bar{A}_{21s}}{6 h_1^2}
-\frac{\bar{A}_{g21s}}{6 h_2^2}\right. \nonumber \\
&+& \left. c_1(h_1-\bar{d}_1) +c_2(h_2-\bar{d}_2) +\bar{S}_1 \exp{(- h_1)}+
\bar{S}_2 \exp{(h_1-h_2)}\right]\,dx,
\label{ENRG_SCL}
\end{eqnarray}
involves the scaled Hamaker constants
$\bar{A}_{ijkl} = \left[(d_2-d_1)/l\right]^2 A_{ijkl}/|A_{12g}|$,
spreading coefficients
$\bar{S}_i = 2 \pi \left[(d_2-d_1)\right]^2 S_i \exp{(d_0/l)} / |A_{12g}|$, and
mean layer thicknesses $\bar{d}_i = d_i/l$.
The $c_i$ are Lagrange 
multipliers that ensure mass conservation for the two liquids.
The corresponding energy scale is $|A_{12g}|/2\pi(d_2-d_1)^2$ and
the ratios of the mean layer thicknesses, surface tensions and viscosities are 
$d = d_2/d_1$, $\sigma = \sigma_2 / \sigma_1$
and $\mu=\mu_2/\mu_1$, respectively. 
Further on, we denote the scaled variables using the 
same symbols as before, i.e.\
the scaled mean thicknesses are given by
$d_i$ and the local thicknesses by $h_i$.
The non-dimensional mobility matrices are obtained from  
Eqs.\,(\ref{MOB}) and (\ref{TANG}) by 
dropping the factor $1/ \mu_1$ and replacing $\beta$ by $\beta / l$.

\section{Linear stability}
\label{lin_stab}
\subsection{General stability of flat films}
\label{gen_stab}
We start the analysis of our model for two-layer films
by discussing the linear 
stability of flat films with $h_1 (x) = d_1$ and $h_2 (x) = d_2$. 
Eqs.\,(\ref{EVOL_PHYS}) are linearized in $\epsilon\ll1$ for small amplitude 
disturbances $\epsilon\chi_i \exp{(\gamma t)} \exp(kx)$ for $i=1,2$
where $k$, $\gamma$ and ${\bm \chi}=(\chi,1)$ 
are the wave number, growth rate and amplitudes of the disturbance, 
respectively. 
The dispersion relation $\gamma(k)$ is obtained by solving the resulting
eigenvalue problem 
$({\bm J}-\gamma {\bm I}) {\bm \chi} = 0$. 

For the isothermal case ($\partial_x \sigma_{12} =\partial_x \sigma_{2}=0$ in
Eqs.\,(\ref{EVOL_PHYS})), 
the corresponding $non$-symmetric Jacobi matrix ${\bm J}$ is given by 
${\bm J} = -k^2 {\bm Q}\cdot{\bm E}$, where ${\bm Q}$ is the scaled mobility matrix.
${\bm E}$ is the energy matrix
\begin{equation}
{\bm E} = \left(
\begin{array}{cc}
\frac{\partial^2 f}{\partial h_1^2} + k^2& 
\frac{\partial^2 f}{\partial h_1 \partial h_2} \\ 
 \frac{\partial^2 f}{\partial h_1 \partial h_2} & 
\frac{\partial^2 f}{\partial h_2^2} +\sigma k^2
\end{array}
\right),
\label{ENRG_MATRIX}
\end{equation}
where $f(h_1,h_2)$ is the local part of the energy density from 
Eq.\,(\ref{ENRG_SCL}). 
This yields
\begin{equation}
\gamma = \frac{\text{Tr}}{2} \pm \sqrt{\frac{\text{Tr}^2}{4} -{\rm Det}},
\label{GR_RATE}
\end{equation}
where ${\rm Tr} = -k^2 [2 Q_{12} E_{12} + Q_{11} E_{11}  + 
Q_{22} E_{22}]$
and 
${\rm Det} = k^4 \det {\bm Q} \det {\bm E}$ are the trace and the determinant of 
${\bm J}$.
Since $\det {\bm Q} \neq 0$ the eigenvalue problem can be written as 
the generalized eigenvalue problem 
$(k^2 {\bm E} + \gamma {\bm Q^{-1}}) {\bm \chi} = 0$.
Because ${\bm E}$ and ${\bm Q^{-1}}$ are both symmetric and ${\bm Q^{-1}}$ is
positive definite one can deduce that all eigenvalues
$\gamma$ are real \cite{MSV87} as expected for a variational problem.
In the non-isothermal case, the Jacobi matrix is given by 
${\bm J} = -k^2({\bm Q}\cdot{\bm E} -{\bm D}\cdot{\bm \Gamma})$, 
where ${\bm \Gamma}$ is a
scaled matrix of coefficients for the Marangoni terms. It is defined in 
Section~\ref{appendix}.
Neither the matrix ${\bm D}$ nor ${\bm \Gamma}$ are symmetric.
This leads to (in general) complex eigenvalues indicating the possibility of 
oscillating motion in the non-isothermal case \cite{NeSi90,NeSi97}.

Going back to the isothermal case, inspection of the generalized eigenvalue
problem shows that the stability
threshold is completely determined by the eigenvalues of ${\bm E}$.
Since the surface tension terms are always positive, 
the onset of the instability is always found for $k=0$, i.e.\ the system is 
linearly stable, independently of the wavelength of the disturbance, for
\begin{eqnarray}
\det {\bm E} > 0 \quad\text{and}\quad E_{11}> 0 
\quad\text{at}\quad k=0.
\label{STAB}
\end{eqnarray}
An instability sets in
if at least one of the conditions (\ref{STAB}) is violated.
Then the flat two-layer film is unstable to disturbances with $k$ larger zero and 
smaller than a cutoff wavenumber 
\begin{equation}
k_c^2 = -\frac{1}{2} \left( \frac{\partial^2 f}{\partial h_1^2}
+\frac{1}{\sigma}\frac{\partial^2 f}{\partial h_2^2 } \right) \pm 
\sqrt{\frac{1}{4} \left(\frac{\partial^2 f}{\partial h_1^2}
-\frac{1}{\sigma}\frac{\partial^2 f}{\partial h_2^2 } \right)^2 + 
\frac{1}{\sigma}\left( \frac{\partial^2 f}{\partial h_1 \partial h_2} \right)^2}
\label{CUT_OFF}
\end{equation}
determined by the condition $\det {\bm E}(k_c) = 0$.

Fig.\,\ref{fig2} shows a schematic stability diagram in the
plane $(E_{11},E_{22})$.
The stability threshold $E_{11} E_{22} = E_{12}^2,~E_{11} >0$ 
is a hyperbola, represented by the solid line.
The unstable region below and left of that line is divided by a second
hyperbola into a two-mode and a one-mode region. 
In the two-mode region both growth rates given by Eq.\,(\ref{GR_RATE}) 
are positive for
$k$ smaller then the respective cut-off $k_c$.
In the one-mode region only one $\gamma$ is positive for $k<k_c$.
Fixing all other system parameters, $\det {\bm E}(k)$ is
determined by $k$. If at $k=0$ the system is in the two-mode region, 
then by increasing $k$ one passes two times
a line $\det {\bm E} = 0$, as indicated by the dashed arrow in Fig.\,\ref{fig2}.
At each crossing a growth rates (Eq.\,(\ref{GR_RATE}))
becomes negative, i.e.\ a mode is stabilized.
If  at $k=0$ the system is in the one-mode region 
the line $\det {\bm E} = 0$ is crossed only once (dot-dashed line). 
\subsection{Long-range interaction only}
\label{lr_int}
As detailed above the stabilizing short-range interaction
only becomes important if at least one layer thickness is locally comparable to the
interaction length $l$, i.e.\ if a layer becomes thinner than about $10$\,nm.
Therefore, to study the linear stability of thicker layers one can neglect 
the short-range terms in Eq.\,(\ref{ENRG_SCL}). 
In this case, Eq.\,(\ref{STAB}) is used to study the role of the Hamaker 
constants Eq.\,(\ref{HAM}) in the linear evolution of the system.
First we note that the Hamaker constants are coupled through the refractive
indices of the media $n_i$. This allows only for selected 
combinations of signs of the $A_{ijk}$ and $A_{ijkl}$ as given in
Table~\ref{Ham_Table}. 

For fixed Hamaker constants, i.e.\ fixed combination of materials, 
$\det {\bm E}_0 = \det {\bm E} (k=0)$ 
is a function of the ratio $d$ of the layer thicknesses only.
Using Table~\ref{Ham_Table} one can
show that for positive $\partial^2 f / \partial h_i^2$ the equation 
$\det {\bm E}_0 (d) = 0$ can only have the solution $d=1$, i.e.\ $d_2-d_1=0$. 
This means that only for vanishing upper layer the system can be on the stability threshold.
In consequence the stability threshold
can $not$ be crossed by solely changing the ratio of layer thicknesses.
This was analyzed in Ref.\,\onlinecite{PBMT04} for a variety of experimentally 
studied systems. Increasing the ratio $d$ from one the system remains either
completely in the unstable or in the stable region.

To compare the stability behaviour of two-layer and one-layer films we
introduce two effective one-layer films as follows.
In (1) we assume the lower layer to be solid, i.e.\ we regard 
the upper layer as a one-layer film  on a coated substrate.
In (2) we assume the upper layer to be rigid but deformable by bending. 
The lower liquid layer corresponds then to
a one-layer film on a solid bulk substrate. 
In case (1) the one-layer liquid film is unstable if the second derivative 
of the energy with respect to the film thickness $h_2$ is negative, 
$\partial^2 f/ h_2^2  < 0$.
The stability threshold at $\partial^2 f/ h_2^2  = 0$ can be crossed by changing
the layer thickness $h_2-h_1$ or the thickness of the coating layer $h_1$.
This was demonstrated in Refs.\,\onlinecite{MSS03} and \onlinecite{SHJ01} 
for a PS film on Si wafers covered with a $1.6$ nm thick SiO layer.
In case (2) the one-layer liquid film is unstable for $\partial^2 f /h_1^2 < 0$.
It can also be destabilized by changing the layer thicknesses, as was
shown in Ref.\,\onlinecite{DRSS98} for a rigid PS layer on top of a liquid PDMS 
layer on a Si substrate.

Comparing the stability thresholds for 
the two effective one-layer systems to the stability diagram in
Fig.\,\ref{fig2} shows that the 
stability threshold of the two-layer system lies in the region where
both effective one-layer systems are stable. 
This indicates that a two-layer system is always less stable than corresponding
effective one-layer systems. 
\subsection{Different instability modes}
\label{inst_modes}
The stability threshold can be studied in rather general terms as was done
above because its  main features do not depend on surface tensions or 
viscosities. However, this is not the case for the characteristics of the
instability like mode type, growth rates or dominant wave length.
To discuss these we focus in the
following on selected two-layer films studied experimentally \cite{MSS03,LPHK96,Sfer97}.
We consider various combinations of layers 
of polystyrene (PS), poly(methylmethacrylate) (PMMA) and 
poly(dimethylsiloxane) (PDMS) on a silicon (Si) or on a silicon-oxide (SiO)
substrate.
The Hamaker constants for different combinations are
calculated using Eq.\,(\ref{HAM}) and given in Table~\ref{Ham}.

The linear instability of a two-layer film has two different modes. It can be of 
zigzag or varicose type. For the former the deflections of the two
interfaces are in phase whereas for the latter they are in anti-phase. 
For special parameter values one can also find a mixed type, where modes are present 
because they have equal fastest growth rates \cite{PBMT04}. 

The model studied in 
Ref.\,\onlinecite{BMR93} assumes a thick lower layer thereby neglecting 
the interaction between the substrate and the 
liquid-liquid and the liquid-gas interface. In this case only the
varicose mode can be unstable.
In the general case, however, also the zigzag mode can become unstable.
Both modes are normally asymmetric, i.e.\ the deflection amplitudes of 
the two interfaces differ.
We characterize this asymmetry by $\phi = \chi/(1+\chi^2)$.
Negative (positive) $\phi$ corresponds to varicose (zigzag) modes.
The value $|\phi| = 1/2$ represents the symmetric case, whereas $\phi = 0$ corresponds to
maximal asymetry, i.e.\ one of the interfaces is flat.
The asymmetry increases with the ratio of the surface tensions $\sigma$. 
Note, that the dispersion relation and the type of the dominant mode
depend on $\sigma$ and $\mu$, whereas the stability diagram Fig.\,{\ref{fig2}}
{\it does not}.

The two mode types are plotted in Fig.\,{\ref{fig3}} for a Si/PMMA/PS/air 
system for a fixed value of $d$ and different ratios of viscosities $\mu$. 
The dispersion relations $\gamma(k)$ are shown together with the 
corresponding $\phi$.
We show here that the type of the dominant mode can be changed 
by varying $d$ or $\mu$. It can also change in dependence of the ratio of the
interfacial tensions $\sigma$ as studied in Ref.\,\onlinecite{PBMT04}.

Strictly speaking, the concept of the mode type characterized by
$\phi$ is only valid in the linear stage of the evolution. 
However, to discuss morphology changes we generalize this concept to 
nonlinear thickness profiles $h_i (x)$. 
We define a generalized mode or solution type by the integral
\begin{equation}
\phi_{\text{int}} = \frac{1}{L} \int \frac{(h_1-d_1)(h_2-d_2)}{[(h_1-d_1)^2+(h_2-d_2)^2]}\,dx,
\label{int_phi}
\end{equation}
taken over the domain length $L$. In many cases the sign of the product
$(h_1 -d_1)(h_2 - d_2)$ does not depend on $x$ allowing to 'read' the mode type
directly from the plots of the layer profiles. For small deflection amplitudes
Eq.\,(\ref{int_phi}) gives again the linear mode type defined above. In the
following we use the notion 'mode-type' in the linear and in the nonlinear regime.

\subsection{Limiting cases}
\label{lim_cases}
For general $d_i$ the radical in Eq.\,(\ref{GR_RATE}) does not allow to give
an analytic expression for the wave number $k_m$ 
and the characteristic growth time $\tau_m = 1/\gamma_{\rm m}$ of the fastest growing mode. 
Nevertheless, one can derive asymptotic expressions for $k_m$ and $\tau_m$ in the two 
important limiting cases of (1) small thickness of the upper layer 
$d_2-d_1 \ll d_2$ 
and (2)  small thickness of the lower layer $d_1 \ll d_2$.
First consider case (1), which corresponds to a liquid film (the upper layer)
on a liquid substrate (the very thick lower layer). 
The dimensional $k_m$ and $\tau_m$  are then given by 
\begin{eqnarray} 
k_m =\frac{1}{(d_2-d_1)^2}\sqrt{\frac{
|A_{12g}|}{4 \pi \sigma_{\rm eff}}} \nonumber \\
\tau_m  = \frac{16 (2 \pi)^2 \sigma_{\rm eff} \mu_1 
(d_2-d_1)^6}{d_1 A_{12g}^2},
\label{1LIM}
\end{eqnarray}
with $\sigma_{\rm eff} = \sigma_1 \sigma_2 /(\sigma_1 +\sigma_2)$.
Note that all variables are in their dimensional form.

Interestingly, the growth time $\tau_m$ depends only on the viscosity 
of the lower layer $\mu_1$ and does not depend on $\mu_2$.
This can be explained by the fact that the flow in the lower layer which
is related to $\mu_1$, is much larger 
than that in the upper one \cite{BMR93}.
At constant thickness of the lower layer, $\tau_m$ is proportional
to $(d_2-d_1)^6$, i.e.\ a liquid film on a bulk liquid substrate
evolves faster than the same film on a solid substrate 
$({\rm growth}~{\rm time} \sim (d_2-d_1)^5)$ and even faster than the 
same film 
on a solid substrate with slippage $( {\rm growth}~{\rm time} \sim 
(d_2-d_1)^5/\left[1 + 3 \beta/(d_2-d_1)\right]$).

In case (2), which corresponds to a liquid film (the lower layer) on a solid substrate 
below the other liquid (the very thick upper layer), the dimensional $k_m$ 
and $\tau_m$  are given by
\begin{eqnarray} 
k_m =\frac{1}{d_1^2}\sqrt{\frac{A_{21s}}{4 \pi \sigma_1}} 
\nonumber \\
\tau_m  = \frac{12 (2 \pi)^2 \sigma_1 \mu_1 
d_1^5}{A_{21s}^2}.
\label{2LIM}
\end{eqnarray}
Note that in case (2) $k_m$ and $\tau_m$ coincide with 
$k_{\rm low}$ and $\tau_{\rm low}$, respectively, the characteristics
of the dominant mode of the instability of a liquid film below a bulk liquid 
calculated using one-layer theory. A discussion of this geometry 
for a Rayleigh-Taylor instability can be found in Refs.\,\onlinecite{YiHi89}
and \onlinecite{YiHi91}.
\subsection{Long-range apolar and short-range polar interactions}
\label{lr_sr_int}
The stability analysis based only on long-range interactions
becomes incorrect for layer thicknesses in the range of the
interaction length $l$ of short-range interactions.
Practically, the latter become important (well) below 
$10$\,nm layer thickness. In contrast to the result for the exclusive action of long-range
van der Waals forces, in the regime where both, short- and long-range interactions, are important
the stability threshold can be crossed by changing the layer
thicknesses $d_i$. 
Fig.\,\ref{fig4} presents a selection of qualitatively different 
stability diagrams in the plane spanned by the layer
thicknesses obtained when varying the strength of the short-range interaction 
for a fixed long-range interaction.

By changing the short-range part of the spreading coefficient $S_1$ and $S_2$
one finds seven topologically different types of such diagrams. These types correspond
to regions in the $(S_1,S_2)$ plane as indicated in Fig.\,\ref{fig5}.
In the absolute unstable region bounded on the right 
by $(S_1)_{\rm min} = (e/4)^4 A_{21s}/|A_{12g}|$ 
and above by $(S_2)_{\rm min} =(e/4)^4$ the system can not be stabilized 
by changing $d_1$ or $d_2$. 
Only if at least one of the two $S_i$  is larger than
the corresponding critical value (a) stable region(s) can be found 
in the $(d_1,d_2-d_1)$ plane (see Fig.\,\ref{fig4}).
For $S_1 > (S_1)_{\rm min}$ a stable region exists that extends towards infinite 
$(d_2-d_1)$,  as shown in Figs.\,\ref{fig4}(a), (b) and (d). 
Thereby, for large $(d_2-d_1)$  the system is stable for
$(d_1)_{\rm min}< d_1 < (d_1)_{\rm max}$, where $(d_1)_{\rm max}$ and
$(d_1)_{\rm min}$ are the solutions of the equation 
$A_{21s}/|A_{12g}| = S_1 x^4 \exp{(-x)}$.
Similarly, for $S_2 > (S_2)_{\rm min}$ a stable region exists that 
extends towards infinite $d_1$, 
as in Figs.\,\ref{fig4}(a) to (d).  For large $d_1$ the system is stable for
$(d_2-d_1)_{\rm min}< d_2 < (d_2-d_1)_{\rm max}$,
where $(d_2-d_1)_{\rm max}$ and
$(d_2-d_1)_{\rm min}$ are the solutions of the equation 
$1 = S_2 x^4 \exp{(-x)}$. 
In the gray shaded triangle at the center of
Fig.\,\ref{fig5} an additional bounded stable region exists in the 
$(d_1,~d_2-d_1)$ plane (see Figs.\,\ref{fig4}(b) and (c)).
Combining the different conditions gives the following seven types of
stability diagrams.
\begin{itemize}
\item[\bf I:] The stable region is continuous and extends in respective
stripes towards infinite $d_1$ and $d_2-d_1$ (Fig.\,\ref{fig4}(a)).
\item[\bf II:] There exist two separated stable regions, one extending towards infinite
$d_1$ and the other one towards infinite $d_2-d_1$ (Fig.\,\ref{fig4}(d)).
\item[\bf III:] Similar to Type II but with an additional bounded stable region 
(Fig.\,\ref{fig4}(b)).
\item[\bf IV:] A bounded stable region exists together with an unbounded
region extending towards infinite $d_1$ (Fig.\,\ref{fig4}(c)).
\item[\bf V:] Similar to type IV but with the unbounded region extending towards
infinite $(d_2-d_1)$ (not shown).
\item[\bf VI:] Only one stable region exists extending towards infinite
$d_1$ (not shown).
\item[\bf VII:] Similar to type VI but with the unbounded region extending towards
infinite $(d_2-d_1)$ (not shown). 
\end{itemize}
Further on we will focus our attention on the stability diagram of type I.

\section{Non-linear behaviour}
\label{nuss}
\subsection{Stationary solutions as extrema of the Lyapunov functional}
\label{nu_sol}
To find periodic stationary solutions of the scaled Eqs.\,(\ref{EVOL_PHYS}),
the time derivatives $\partial_t h_i$ are set to zero.
Integration yields 
\begin{eqnarray}
Q_{11}\, \partial_x \left( \frac{\delta F}{\delta h_1} \right)+
Q_{12}\, \partial_x \left( \frac{\delta F}{\delta h_2} \right) &=& C_1 
\nonumber \\
Q_{21}\, \partial_x \left( \frac{\delta F}{\delta h_1} \right)+
Q_{22}\, \partial_x \left( \frac{\delta F}{\delta h_2} \right) &=& C_2,
\label{STAT}
\end{eqnarray}
where the $C_i$ are constants and $F$ is given by  
Eq.\,(\ref{ENRG_SCL}).
Note that the left hand sides of Eqs.\,(\ref{STAT}) represent the flow in the lower 
layer and the total flow, respectively. For a stationary state both flows
are zero, i.e.\ the $C_1=C_2=0$. 
Because the mobility matrix ${\bm Q}$ is non-singular, 
one concludes from  Eqs.\,(\ref{STAT}) that the stationary states
of the  Eqs.\,(\ref{EVOL_PHYS}) are the extrema of the Lyapunov functional
$F$, i.e.\ they are solutions of
\begin{eqnarray}
-\partial_{xx} h_1 +\frac{\partial f}{\partial h_1} &=& c_1 \nonumber \\
-\sigma \partial_{xx} h_2 +\frac{\partial f}{\partial h_2}  &=& c_2,
\label{STAT_1}
\end{eqnarray}
where $f$ denotes the local part of Eq.\,(\ref{ENRG_SCL}) and the constants
$c_i$ correspond to the Lagrangian multipliers introduced in Section\,\ref{deriv}.
To obtain a finite amplitude solution for given mean 
thicknesses we use continuation techniques \cite{DKK91,DKK91b,Doedel97}.
We start with analytically known stationary periodic small-amplitude profiles,
which correspond to the linear eigenfunctions for the critical wave number 
$k_c$.
By continuation we follow the family of solutions changing the period $L$.
We characterize the solutions by the deflection amplitudes $A_1$ and
$A_2$, the energy $E$, the norm $L_2 = (1/L) \int[ (h_1-d_1)^2+(h_2-d_2)^2]\,dx$ and
the integral mode type $\phi_{int}$.
To determine the stability of the stationary solutions $h_i(x)$, 
we add small perturbations $\delta h_i (x) \sim exp{(\beta t)}$ to both 
interfaces $h_i(x)$, linearize
the full time-dependent evolution equations (\ref{EVOL_PHYS}) around $h_i(x)$ and
solve the obtained eigenvalue problem 
${\bm L}(h_i, \partial_x h_i, \partial_x ) {\bm \delta \bm h(x)} = 
\beta {\bm \delta \bm h(x)}$ 
for the linear operator ${\bm L}$ after discretizing it in space. 
The sign of the largest eigenvalue $\beta$ determines the stability
of the stationary solution.
Note that due to the translational invariance of the evolution equations
(\ref{EVOL_PHYS}), there exists always a symmetry mode 
$\delta h_i (x) = \partial_x h_i(x)$ with the eigenvalue $\beta = 0$.
\subsection{Mode type transitions}
\label{mt_trans}
The type of the dominant instability mode calculated above 
by linear stability may not persist in the course of the nonlinear evolution. 
Possible mode type changes may have a dramatic effect on the (observable) 
overall morphology of the film. We investigate these changes by studying both,
the evolution in time of the film profiles and the stationary solutions obtained by
continuation. 

The evolution in time is obtained by numerical simulations of the scaled 
coupled evolution equations Eqs.\,(\ref{EVOL_PHYS}) in a one-dimensional 
periodic domain. Both, semi-implicit pseudo-spectral and
explicit time integration schemes are used. 
Initial conditions consist of flat layers with
an imposed noise of amplitude 0.001. 

\subsubsection{Transition via branch switching}
\label{trans_branch}
First the time evolution of an initially flat film is 
studied for parameters as in Fig.\,\ref{fig3}\,(a) using a domain size 
equal to four times the fastest growing wave length $\lambda_m$.
A time sequence of snap shots is shown in Fig.\,\ref{fig6}.
In the early stage of the evolution a varicose mode develops $(t = 8.1)$
as expected from the linear analysis. Then in a sub-domain of size 
$\lambda_m$ the deflection amplitudes increase dramatically
accompanied by a morphological change towards a zigzag type profile
($t = 9.7$). This is further illustrated by the dependence of the 
integral mode-type (Eq.\,\ref{int_phi}) on time given in Fig.\,\ref{fig7}(b).
Further on, the length of the zigzag part increases slightly, and coarsening sets
in resulting in the disappearance of one varicose-type drop ($t = 16.4$).
Next one of the remaining drops increases it amplitude and flips to a 
zigzag type hole ($t=20.6$).
Finally the last remaining varicose-type drop disappears $(t\approx28)$, and
the system approaches a stationary (but not stable) state.
The evolution of the relative energy of the profile in time is given in
Fig.\,\ref{fig7}(a). It is seen very clearly that the phases of very slow 
evolution correspond to solutions that are close to stationary solutions. This
results from the fact that the (unstable) stationary solutions
form saddle points in function space that are approached along their stable
manifolds and subsequently repel the system along their unstable manifolds
(for a discussion see Ref.\,\onlinecite{TBBB03})
The evolution stops after a further coarsening step, when the period becomes 
equal to the system size (not shown).

To explain the observed mode-type change, we study the stationary solutions of
the evolution equations Eqs.\,(\ref{EVOL_PHYS}).   
We find a family of solutions starting at the subcritical primary bifurcation,
then turning three times at saddle-node bifurcations (folds) and going towards
infinite periods (see Fig.\,\ref{fig8}(a)).
A stability analysis using the solution period as the period of the
disturbance (thereby excluding coarsening modes) 
shows that two branches are stable (solid lines) 
and two are unstable (dashed lines).
Along the first unstable branch, which starts at $L_c$ and ends at 
the first fold at $L\approx 60$ the energy 
$E$ is always larger than the one of the flat film $E_0$
(Fig.\,\ref{fig8}(b)), and it increases with decreasing period.
This subcritical branch corresponds to nucleation solutions that have to be overcome 
to break the film in parts smaller than $L_c$ (see Ref.\,\onlinecite{TNPV02} for a discussion of this
type of solutions for a one-layer system).
The first stable branch starts at the first fold at $L \approx 59$ and ends at the 
second fold at $L \approx 116$. Its relative energy decreases monotonically with increasing
period. Mostly it is energetically preferable to the flat film.
The second unstable branch (between the second fold at $L \approx 116$ 
and the third fold at $L \approx 79$) turns back towards smaller periods.
The second stable branch starts at the third fold and goes towards
infinite periods. Its energy decreases rapidly from values even 
above the flat film to values below the ones of the first stable branch.
The energy of the second unstable branch is always larger than the energies of
both stable branches. This indicates that it corresponds to nucleation
solutions, or critical solutions that have to be overcome to switch between
the two stable branches.
Along the second unstable branch, the mode-type changes from varicose to
zigzag (Fig.\,\ref{fig8}\,(d)) explaining the non-trivial behavior observed in
the time evolution shown in Fig.\,\ref{fig6}. There are two stable solutions
with a period equal to the dominant linear wave length 
$(\lambda_m \approx 108)$ (see Fig.\,\ref{fig9}). 
The one of higher energy that is approached first
in the time evolution is of varicose type whereas the one of lower energy that
the system switches to is of zigzag type (cp.\ Fig.\,\ref{fig8}(b)).
A transition between the two solutions is accompanied by a strong increase of 
the amplitude $A_1$ (see Fig.\,\ref{fig8}(a)).

\subsubsection{Transition via coarsening}
\label{trans_coars}
A mode-type change is not always connected to a transition
between different branches of stationary solutions. Also coarsening along one 
branch may lead to such a change if the mode-type varies along the branch.
To demonstrate this, we simulate the time evolution using parameters as in
Fig.\,\ref{fig3}(b).
A time sequence of profiles and the corresponding
evolution of the relative energy and the integral mode-type are shown in
Figs.\,\ref{fig10} and \ref{fig11}, respectively.
Early in the evolution the layer profiles
represent a zigzag mode $(t = 6.0)$ corresponding to the linear results 
(Fig.\,\ref{fig3}(b)). 
Then, within the very short period of time from $t=6.0$ to $10.8$,
nonlinear effects result in a first change towards a varicose
type profile, as shown in the inset of Fig.\,\ref{fig11}(b). 
Then the system approaches the branch of stationary solutions. As a result the
evolution slows down and the pattern begins to coarse.
With ongoing coarsening $(t >10.8)$ the size of the droplets increases
$(t=135,~t=461)$ and at very late times $(t>490)$ the mode type 
changes back to zigzag type (Fig.\,\ref{fig11}(b)).
Here, the amplitudes of the interfaces do not change dramatically, as was the case 
in Section~\ref{trans_branch}. In this sense the transition is continuous.
The characteristics of the corresponding stationary solutions are shown 
in Fig.\,\ref{fig12}.
The primary bifurcation is again subcritical (Fig.\,\ref{fig12}(a)) the
solution family continues towards smaller periods until turning at a
saddle-node bifurcation (fold) and heading towards infinite periods.
The subcritical branch is unstable with energies higher than the
energy of the flat film (Fig.\,\ref{fig12}(b)). The second branch starting
at the fold $(L \approx 26)$ consists of solutions whose energy 
decreases monotonically with increasing period. They 
are stable to disturbances of identical period but unstable to coarsening
modes. 
Fig.\,\ref{fig12}\,(d) shows that the solution with the period equal to
$\lambda_m = 50.81$ is of varicose type. The corresponding layer profile 
is shown in Fig.\,\ref{fig13}(a) together with the profile after the first
coarsening step.
When the period becomes 
larger than $94.2$ , the solution changes to zigzag type (Fig.\,\ref{fig12}(d)) 
as shown in Fig.\,\ref{fig13}(b). 
This explains the mode-type change found in the time evolution (Fig.\,\ref{fig10}).

Here we have restricted our attention to a parameter set corresponding to 
region I of Fig.\,\ref{fig5}, i.e.\ corresponding to the stability diagram
shown in Fig.\,\ref{fig4}\,(a). The existence of a stable branch 
of stationary solutions which continues towards infinite period implies
that the rupture of the two layers is completely avoided by the short-range
repulsion. 
However, this may not be the case for parameter ranges belonging to the other
types of stability diagrams. A detailed analysis of the stationary solutions 
for all types will be done elsewhere.

\subsection{Large-period stationary solutions}
\label{lp_sol}
The stability of the numerical code, used to solve the evolution 
equations Eqs.\,(\ref{EVOL_PHYS}), requires a very small time step $t =0.00001$.
As a result it takes very long even 
to reach the final stationary solution in a system of size $4\lambda_m$ using
$256$ grid points. 
It is not feasible at the moment to study many coarsening steps in this way. 
However, one can rely on continuation techniques that use an adaptive spatial grid
along the continuation path \cite{Doedel97}
to obtain stationary solutions 
of arbitrarily large periods that correspond to solutions that would be
obtained 
in a time evolution at very late times.
We show in Fig.\,\ref{fig14} possible large-period stationary solutions 
for different physical systems that are investigated experimentally. 
One finds qualitatively different  morphologies like a drop of the lower
liquid 'looking through' a nearly flat film of the upper layer for a 
Si/PMMA/PS/air system. Note however that also the upper layer is continuous
(due to the stabilizing short-range interaction), i.e.\ also the drop is
covered by a very thin layer of the upper liquid. This is more pronounced 
for a SiO/PS/PDMS/air system. In contrast, for a SiO/PMMA/PS/air system one 
finds a drop of the upper liquid 'swimming' on the lower liquid that however
is attracted towards the base of the drop. These equilibrium solutions are
equivalents of drop configurations studied in Ref.\,\onlinecite{MAP02}
for macroscopic (but smaller than the capilary length) drops. However, here
the mesoscopic contact angles are not given explicitely but result from the 
underlying effective molecular interactions, i.e.\ the short- and long-range
forces used.
\section{Conclusion}
\label{conc}
We have derived coupled non-linear evolution equations
for the profiles of the liquid-liquid and liquid-gas interfaces of a thin two-layer
liquid film heated from below allowing for slip at the substrate.
We have shown that in the isothermal case the evolution equations can be
written in terms of variations of an appropriate
Lyapunov functional $F$ which monotonically decreases in time.
The stability conditions for flat layers have been given in terms of $F$.
We have shown that a two-layer film is less stable
than related effective one-layer films introduced in Section~\ref{lr_int}.
Even if both effective one-layer films are stable the two-layer film may
be unstable if the determinant of the energy matrix $\det {\bm E}$ is
negative. 

We have shown that 
if the Hamaker constants are given by the usual expression (Eq.\,(\ref{HAM})),
i.e.\ they are coupled through the refractive indices,
and no other forces are present, the stability of the flat films with
thicknesses of ($\sim100$\,nm) 
can not be changed by solely changing the layer thicknesses.
Incorporating a stabilizing short-range interaction the stability can be
changed in this way. We have classified the resulting
possible types of stability diagrams in the space of the layer thicknesses
and given a 'phase diagram' in terms of the short-range parts of the spreading
coefficients for the occurence of the different types of stability diagrams.

In general, the linear stability analysis of the flat film has shown that both,
varicose or zigzag mode, may be unstable depending on 
the ratios of the layer thicknesses, viscosities and surface tensions (see
also Ref.\,\onlinecite{PBMT04}). This seems to be in contrast to
Ref.\,\onlinecite{BMR93}. However, the difference arises because 
there it is assumed that the 
lower layer is thick compared to the upper layer neglecting thereby all
interactions with the substrate. Then the zigzag mode is always stable.

The introduction of the stabilizing short-range interaction allows
to study the long-time evolution and stationary layer profiles.
Possible stationary states have then been determined as
extrema of the Lyapunov functional $F$. The resulting bifurcation
diagrams show a rich branch structure that depends strongly on
parameter values. We have focused on one type of stability diagram where
a stable branch going towards infinite period always exists. This implies
the existence of a non-ruptured stationary state in the long-time limit
also in the time evolution.

We have found that the mode type of a profile may change during the 
evolution of an instability on three ways. First, the profile type changes
in the course of the short-time evolution. This is connected to different mode
types found for the dominant linear mode and the stationary solution of equal
period on the solution branch approached first in the time evolution.
It seems that this behaviour is more probable for a subcritical primary
bifurcation. In the case studied here this change is from zigzag to varicose.
In the nonlinear regime the profile can change its type by 
(i) jumping from one to another stable branch and by (ii) coarsening along a 
single stable branch. Combinations of the different ways may also be possible.
We have found that for the parameters considered here both nonlinear
transitions go from varicose towards zigzag type. 
In case (i) the transition occurs without change of the period, but with a
dramatic increase in amplitude of the profile.
In the case (ii) the transition occurs continuously without amplitude
jump because mediated by coarsening a small-period varicose mode turns into a 
large-period zigzag one.

In all examples considered here (except the SiO/PMMA/PS/air system with
$d_1 = 30$, $d_2 = 50$) we have found a zigzag-type solution at large 
periods. For the future it would be very interesting to further analyze
the stationary solutions for a broader range of experimentally interesting
systems like the ones studied in Refs.\,\onlinecite{MSS03,PWHC97}. This should
clarify under which conditions the long-time (or large-period) solutions
are energetically preferable and determine how 'late' the transition
may occur. A systematic analysis of all types of stability diagrams would also
discuss metastability and absolute stability of the flat two-layer films. 
Furthermore, we are very optimistic that the evolution equations
presented here will serve to study the questions of hole growth and possible 
front instabilities  in the dewetting of a liquid layer on a liquid substrate
of finite thickness\cite{FCW95,PWHC97}.
 
\appendix
\section{}
\label{appendix}
To further specify the thermocapillar part of Eqs.\,\ref{EVOL_PHYS} we
rewrite the derivatives $\partial_x \sigma_{12}$ and  $\partial_x
\sigma_{2}$ in terms of the gradients of the local thicknesses $\partial_x
h_i$.
In the long-wave approximation \cite{ODB97} the temperature field is in both 
layers a linear function of the vertical coordinate $z$, i.e.\, $T_i = a_i z+b_i$. 
To determine the coefficients we consider a three-layer geometry 
(Fig.\,\ref{fig15}), i.e.\ we take into account 
the heat conduction in a gas layer of finite thickness $d_g = d_t -d_2$, where $d_t$ is
the distance between the substrate and an upper plate. 
The temperature in the gas layer is $T_g = a_g z +b_g$.
The boundary conditions at both interfaces are continuity of the temperature
field and continuity of the heat flux 
$\kappa_i \partial_z T_i=\kappa_k \partial_z T_k$,
where $\kappa_i$ is the thermal conductivity of the $i$-th layer.
The temperatures at the substrate $T_0$ and at the upper plate $T_t$ are 
constant.
The coefficients $a_i$ and $b_i$ depend on the local thicknesses $h_i$ and
are given by
\begin{eqnarray}
a_g = \frac{\alpha \Delta T }{d_t-h_2 + \frac{\kappa_g}{\kappa_1}h_1 +
  \frac{\kappa_g}{\kappa_2}(h_2-h_1)}
\nonumber \\
\nonumber \\
a_2 = \frac{a_g \kappa_g}{\kappa_2},~~~~~a_1 = \frac{a_g \kappa_g}{\kappa_1} 
\nonumber \\
  \nonumber \\
b_1 = T_0,~~~~b_g = T_t -a_g d_t \nonumber \\
b_2 = a_g \kappa_g h_1 \left(  \frac{1}{\kappa_1} -  \frac{1}{\kappa_2}
  \right) + T_0, \nonumber 
\end{eqnarray}
where $\Delta T = T_0 - T_2$ and 
\begin{eqnarray}
\alpha = \frac{d_2 - d_t - \kappa_g h_1 / \kappa_1 -\kappa_g (d_2-d_1)/
  \kappa_2}{\kappa_g h_1 / \kappa_1 -\kappa_g (d_2-d_1)/ \kappa_2}.
\nonumber 
\end{eqnarray}
Here $T_2$ is the temperature of the liquid-gas interface, when both
interfaces are undeformed, i.e.\ for $h_i = d_i$.
The above formulas allow to determine the derivatives 
\begin{eqnarray}
\partial_x \sigma_{12} = \Gamma_{11} \partial_x h_1 + \Gamma_{12} \partial_x h_2 \\
\nonumber
\partial_x \sigma_{2} = \Gamma_{21} \partial_x h_1 + \Gamma_{22} \partial_x h_2, 
\label{sigmas}
\end{eqnarray}
where the matrix ${\bm \Gamma}$ is determined as follows
\begin{equation}
{\bm \Gamma} = a\left(
\begin{array}{ccc}
\frac{ \kappa_g}{\kappa_1}\frac{d \sigma_1}{ d T} 
b  &~~~&
-\frac{ \kappa_g}{\kappa_1}\frac{d \sigma_1}{ d T}h_1(\frac{\kappa_g}{ \kappa_2} - 1) \\ \\

\frac{d \sigma_2}{d T} (
bc-\frac{\kappa_g}{\kappa_2}h_2b ) & &
 \frac{d \sigma_2}{d T} \{  
\frac{\kappa_g}{\kappa_2} ( d_t+h_1 b )-bh_1 (\frac{\kappa_g}{ \kappa_2}-1) \}
\end{array}
\right).
\label{Gamma}
\end{equation}
Here $a = (\alpha \Delta T)/\left[ d_t-h_2 + 
\frac{\kappa_g}{\kappa_1}h_1 +
  \frac{\kappa_g}{\kappa_2}(h_2-h_1) \right]^2$, $b= \kappa_g (1/\kappa_1 -1/\kappa_2)$  
and 
$c = \{d_t-h_2(1-\frac{\kappa_g}{\kappa_2})\}$.
For the linear normal Marangoni effect $d \sigma_{12} / dT$ and 
$d \sigma_{2} / dT$ are negative and constant. 
The Eqs.\,(\ref{sigmas}) are used in Eqs.\,(\ref{EVOL_PHYS}) to
obtain a closed system of equations for $h_1$ and $h_2$.

\newpage
\begin{table}[ht]
\caption{\label{Ham_Table}
Possible combinations of signs of the different Hamaker constants for given order of the 
refractive indices of the involved medias.
}
\begin{ruledtabular}
\begin{tabular}{cccc}
${\rm refractive}~{\rm indices}$ & $A_{12g}$ &$ A_{21s}$ &$ A_{g21s}$ \\
\hline
$n_s > n_1$, $n_1 < n_2$, $n_2 > n_g$ & + & + & - \\
\hline
$n_s < n_1$, $ n_1 < n_2$, $n_2> n_g$ & + & - & + \\
\hline
$n_s < n_1$, $ n_1 > n_2$, $n_2 > n_g$ & - & + & + \\
\hline
$n_s > n_1$, $n_1 > n_2$, $n_2 > n_g$ & - & - & - \\
\end{tabular}
\end{ruledtabular}
\end{table}

\begin{table}[ht]
\caption{\label{Ham}Hamaker constants for various combinations of polymers.}
\begin{ruledtabular}
\begin{tabular}{cccc}
${\rm System}$ & $A_{12g} \times 10^{-20} {\rm Nm}$ & $A_{21s} \times 10^{-20} {\rm Nm}$ & $A_{g21s} \times 10^{-20} {\rm Nm}$ \\
\hline
${\rm Si/PMMA/PS/air}$ & $1.49$  & $3.8$  & $-23.02$ \\
\hline
${\rm SiO/PMMA/PS/air}$ & $1.49$ & $-0.024$ & $0.15$ \\
\hline
${\rm SiO/PS/PDMS/air}$ & $-1.83$ & $0.42$  & $1.25$ \\
\end{tabular}
\end{ruledtabular}
\end{table}
\newpage
\begin{figure}[H]
\caption{
Sketch of the problem in two dimensions. 
The local thickness of the lower layer is $h_1$, the total local film
thickness is $h_2$. 
}
\label{fig1}
\end{figure}
\begin{figure}[H]
\caption{
The schematic stability diagram for fixed coupling $E_{12}$.
Shown are the stability threshold (solid line) and the boundary between the 
one-mode and the two-mode region (dotted line). Both are given by 
$\det {\bm  E} = 0$ for increasing wave number $k$.
Dashed and dot-dashed arrows represent parametric lines given by 
$(E_{11}(k)$ and $E_{22}(k))$. 
The dashed (dot-dashed) line starts at $k=0$ in the two-mode 
(one-mode) region. 
At an intersection of a line $\det {\bm E} (k=0)$ and a 
parametric line one of the growth rates changes its sign.}
\label{fig2}
\end{figure}
\begin{figure}[H]
\caption{
Shown are the growth rate $\gamma$ (solid lines) and the 
mode type $\phi$ (dashed lines) of the leading eigenmode.
(a) A varicose mode from the one-mode region at $d_1 = 30,~d_2=47$ and 
$\sigma=\mu=1$,
(b) a zigzag mode from the one-mode region at $d_1 = 15,~d_2 = 40$ and 
$\sigma=\mu=1$.  
Panel (c) gives $\gamma$ and $\phi$ for $d_1$ and $d_2$ as in (b) but for 
$\mu=0.1$. 
For convenience we plot in (b) $10 \gamma$  and
 in (a) $20 \gamma$.}
\label{fig3}
\end{figure}
\begin{figure}[H]
\caption{
Different types of stability diagrams in the plane of the layer 
thicknesses $(d_1,d_2-d_1)$,
shown for different strength of the short-range interactions $S_1$ and 
$S_2$ as given in the legends. The shaded parts represent linearly stable regions. The
Hamaker constants are $A_{12g} = 1.49$, $A_{21s} = 3.8$, $A_{g21s} = -23.02$,
corresponding to the Si/PMMA/PS/air system.
Panels (a), (b), (c) and (d) correspond to ranges I, III, IV and II in 
Fig.\,\ref{fig5}, respectively.}
\label{fig4}
\end{figure}
\begin{figure}[H]
\caption{
 Phase diagram in the plane $(S_1,~S_2)$ for the Si/PMMA/PS/air system. 
The absolute stable
  region (hatched rectangle in the lower left corner) is bounded by 
  $(S_1)_{\rm min} = (e/4)^4 A_{21s}/|A_{12g}|$ from the right and by
  $(S_2)_{\rm min} =(e/4)^4$ from above. The unstable region
is divided into seven qualitatively different subregions, 
described in the main text.}
\label{fig5}
\end{figure}
\begin{figure}[H]
\caption{
Snapshots of the time evolution of a Si/PMMA/PS/air system for $d_1 = 30, d_2 =
  47$, $S_1 = S_2 =1, \sigma =1$ and $\mu = 1$ at times as given in the 
legends. The domain length is
$L = 4 \lambda_m$ and time is in units of $1/ \gamma_m$.}
\label{fig6}
\end{figure}
\begin{figure}[H]
\caption{
Evolution in time of (a) the relative energy $E -E_0$ and (b) the integral 
mode-type $\phi_{int}$ (Eq.\,\ref{int_phi}) for parameters as in Fig.~\ref{fig6}.
Time is in units of $1/ \gamma_m$.
In (a) the dashed lines denote the energy levels which correspond to the
stationary solutions with periods $L=\lambda_m$ 
(first and second line from above), $L=4/3\lambda_m$ (third line from above)
and $L=2\lambda_m$ (the lowest line), taken from Fig.\,\ref{fig8}(b).}
\label{fig7}
\end{figure}
\begin{figure}[H]
\caption{
Characterization of the stationary periodic solutions for the 
system of Fig.\,\ref{fig7}. 
Shown are (a) the amplitude of the lower layer $A_1$, (b) the relative energy $E-E_0$,
(c) the norm $L_2$, and (d) the integral mode type $\phi_{int}$
in their dependence on the period $L$.
In (d) the inset shows a zoom of the region marked by the dashed box.}
\label{fig8}
\end{figure}
\begin{figure}[H]
\caption{
The two stationary solutions with period $L = 108.28$ (cp.\ Fig.\,\ref{fig8})
corresponding to the
wave length of the relevant dominant linear mode $\lambda_{max}$.
Shown are (a) the varicose type and (b) the zigzag type
solution from the first and second stable branch, respectively.
}
\label{fig9}
\end{figure}
\begin{figure}[H]
\caption{
Snapshots of the time evolution of a Si/PMMA/PS/air system for $d_1 = 15$, 
$d_2 = 40$, $S_1 = S_2 = 1$, $\sigma =1$ and $\mu = 1$ at times as given in
the legends. 
The domain length is $L=4 \lambda_m$ and time is in units of
$1/ \gamma_m$.}
\label{fig10}
\end{figure}
\begin{figure}[H]
\caption{
Evolution in time of (a) the relative energy $E - E_0$ and (b) the integral
mode-type $\phi_{int}$ (Eq.\,\ref{int_phi}) for parameters as in
Fig.~\ref{fig10}.
The inset in (b) shows the early-time behaviour of the mode-type.
Time is in units of $1/ \gamma_m$.
In (a) the dashed lines denote the energy levels which correspond to the
stationary solutions with periods $L=\lambda_m$, 
$L=4/3\lambda_m$ and $L=2\lambda_m$, taken from Fig.\,\ref{fig12}(b).}
\label{fig11}
\end{figure}
\begin{figure}[H]
\caption{
Characterization of the stationary periodic solutions, for the 
system of Fig.\,\ref{fig10}.
Shown are (a) the amplitude of the upper layer $A_2$, (b) the relative energy $E-E_0$,
(c) the norm $L_2$ and (d) the integral mode type $\phi_{int}$ in their 
dependence on the period $L$.}
\label{fig12}
\end{figure}
\begin{figure}[H]
\caption{
The two stationary solutions with period $L =  50.8$ and $L =  101.6$ (cp.\ Fig.\,\ref{fig12})
corresponding to once and twice the wave length of the dominant linear mode $\lambda_{max}$, respectively.
To symbolize the coarsening process we show in (a) and (b) two and one period(s), respectively.
The $x$ coordinate is in units of $\lambda_{max}$.}
\label{fig13}
\end{figure}
\begin{figure}[H]
\caption{
Large-period (long-time) stationary profiles for (a) a Si/PMMA/PS/air system with
$d_1 =30$, $d_2 =39$, $\lambda_m = 132$ and period $L=26\times\lambda_m$, 
(b) a SiO/PMMA/PS/air with $d_1 =30$, $d_2 =50$, $\lambda_m = 246$ and period 
$L=20\times\lambda_m$,
(c) a Si/PMMA/PS/air system with $d_1 =30$, $d_2 =70$, $\lambda_m = 118$ and
  $L=21\times\lambda_m$  
and (d) a SiO/PS/PDMS/air with $d_1 =30$, $d_2 =70$, $\lambda_m = 557$ and 
$L=10\times\lambda_m$. 
The remaining parameters are $\sigma =1$, $\mu = 1$, and $S_1 =S_2 =1$.}
\label{fig14}
\end{figure}
\begin{figure}[H]
\caption{Sketch of the system with a gas layer of finite thickness.}
\label{fig15}
\end{figure}
\clearpage
\begin{center}
\vspace*{1cm}
\includegraphics[width=0.7\hsize]{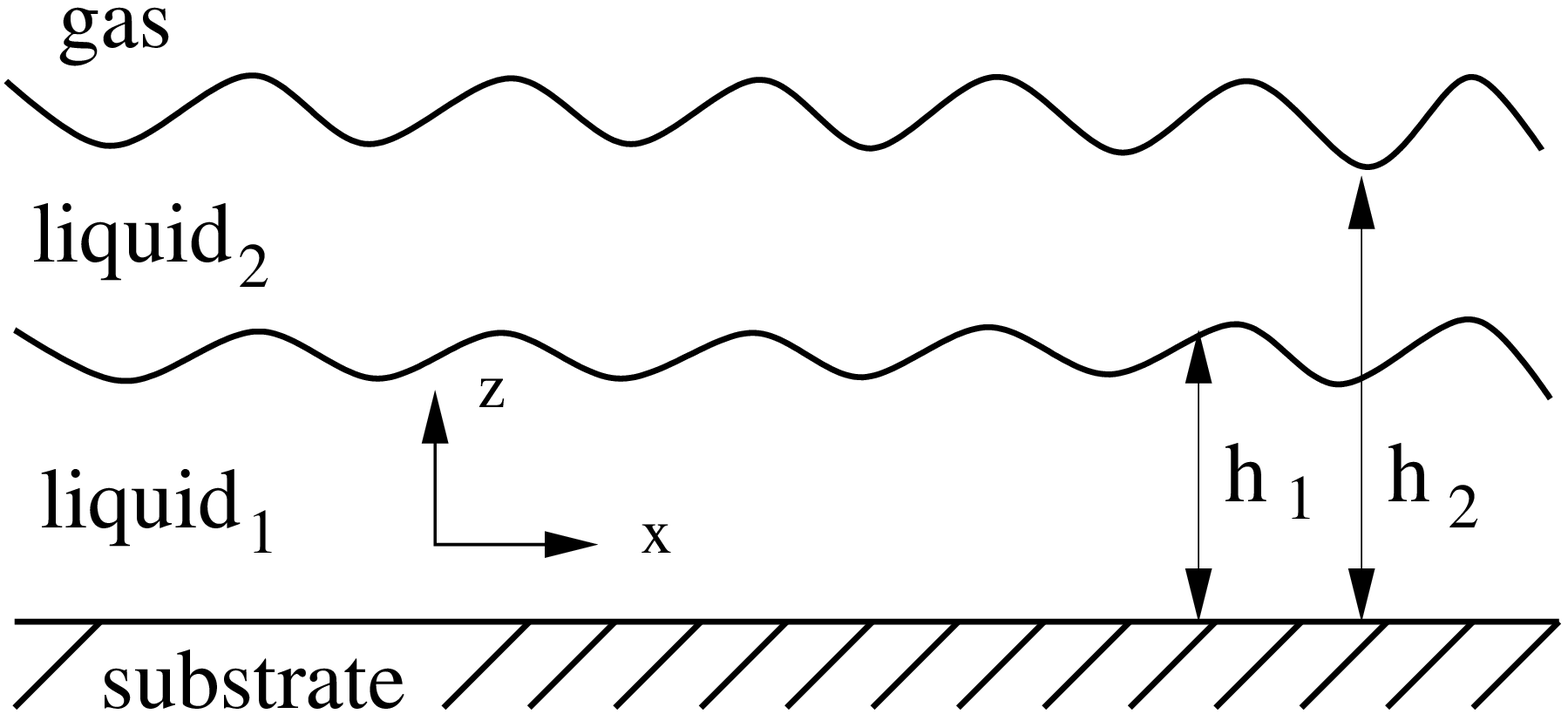}                           
\vspace{4cm}

{\bf\large Fig.\,\ref{fig1}\\
Pototsky et al., J. Chem. Phys.}
\end{center}

\newpage
\begin{center}
\vspace*{1cm}
\includegraphics[width=0.8\hsize]{PBMT05_fig2.eps}                           
\vspace{4cm}

{\bf\large Fig.\,\ref{fig2}\\
Pototsky et al., J. Chem. Phys.}
\end{center}

\newpage
\begin{center}
\vspace*{1cm}
\includegraphics[width=0.7\hsize]{PBMT05_fig3.eps}                           
\vspace{4cm}

{\bf\large Fig.\,\ref{fig3}\\
Pototsky et al., J. Chem. Phys.}
\end{center}

\newpage
\begin{center}
\vspace*{1cm}
\includegraphics[width=0.8\hsize]{PBMT05_fig4.eps}                           
\vspace{4cm}

{\bf\large Fig.\,\ref{fig4}\\
Pototsky et al., J. Chem. Phys.}
\end{center}

\newpage
\begin{center}
\vspace*{1cm}
\includegraphics[width=0.8\hsize]{PBMT05_fig5.eps}                           
\vspace{4cm}

{\bf\large Fig.\,\ref{fig5}\\
Pototsky et al., J. Chem. Phys.}
\end{center}
\newpage
\begin{center}
\vspace*{1cm}
\includegraphics[width=0.8\hsize]{PBMT05_fig6.eps}                           
\vspace{4cm}

{\bf\large Fig.\,\ref{fig6}\\
Pototsky et al., J. Chem. Phys.}
\end{center}
\newpage
\begin{center}
\vspace*{1cm}
\includegraphics[width=0.8\hsize]{PBMT05_fig7.eps}                           
\vspace{4cm}

{\bf\large Fig.\,\ref{fig7}\\
Pototsky et al., J. Chem. Phys.}
\end{center}
\newpage
\begin{center}
\vspace*{1cm}
\includegraphics[width=0.8\hsize]{PBMT05_fig8.eps}                           
\vspace{4cm}

{\bf\large Fig.\,\ref{fig8}\\
Pototsky et al., J. Chem. Phys.}
\end{center}
\newpage
\begin{center}
\vspace*{1cm}
\includegraphics[width=0.6\hsize]{PBMT05_fig9.eps}                           
\vspace{4cm}

{\bf\large Fig.\,\ref{fig9}\\
Pototsky et al., J. Chem. Phys.}
\end{center}
\newpage
\begin{center}
\vspace*{1cm}
\includegraphics[width=0.8\hsize]{PBMT05_fig10.eps}                           
\vspace{4cm}

{\bf\large Fig.\,\ref{fig10}\\
Pototsky et al., J. Chem. Phys.}
\end{center}
\newpage
\begin{center}
\vspace*{1cm}
\includegraphics[width=0.8\hsize]{PBMT05_fig11.eps}                           
\vspace{4cm}

{\bf\large Fig.\,\ref{fig11}\\
Pototsky et al., J. Chem. Phys.}
\end{center}
\newpage
\begin{center}
\vspace*{1cm}
\includegraphics[width=0.8\hsize]{PBMT05_fig12.eps}                           
\vspace{4cm}

{\bf\large Fig.\,\ref{fig12}\\
Pototsky et al., J. Chem. Phys.}
\end{center}
\newpage
\begin{center}
\vspace*{1cm}
\includegraphics[width=0.6\hsize]{PBMT05_fig13.eps}                           
\vspace{4cm}

{\bf\large Fig.\,\ref{fig13}\\
Pototsky et al., J. Chem. Phys.}
\end{center}
\newpage
\begin{center}
\vspace*{1cm}
\includegraphics[width=0.8\hsize]{PBMT05_fig14.eps}                           
\vspace{4cm}

{\bf\large Fig.\,\ref{fig14}\\
Pototsky et al., J. Chem. Phys.}
\end{center}
\newpage

\begin{center}
\vspace*{1cm}
\includegraphics[width=0.7\hsize]{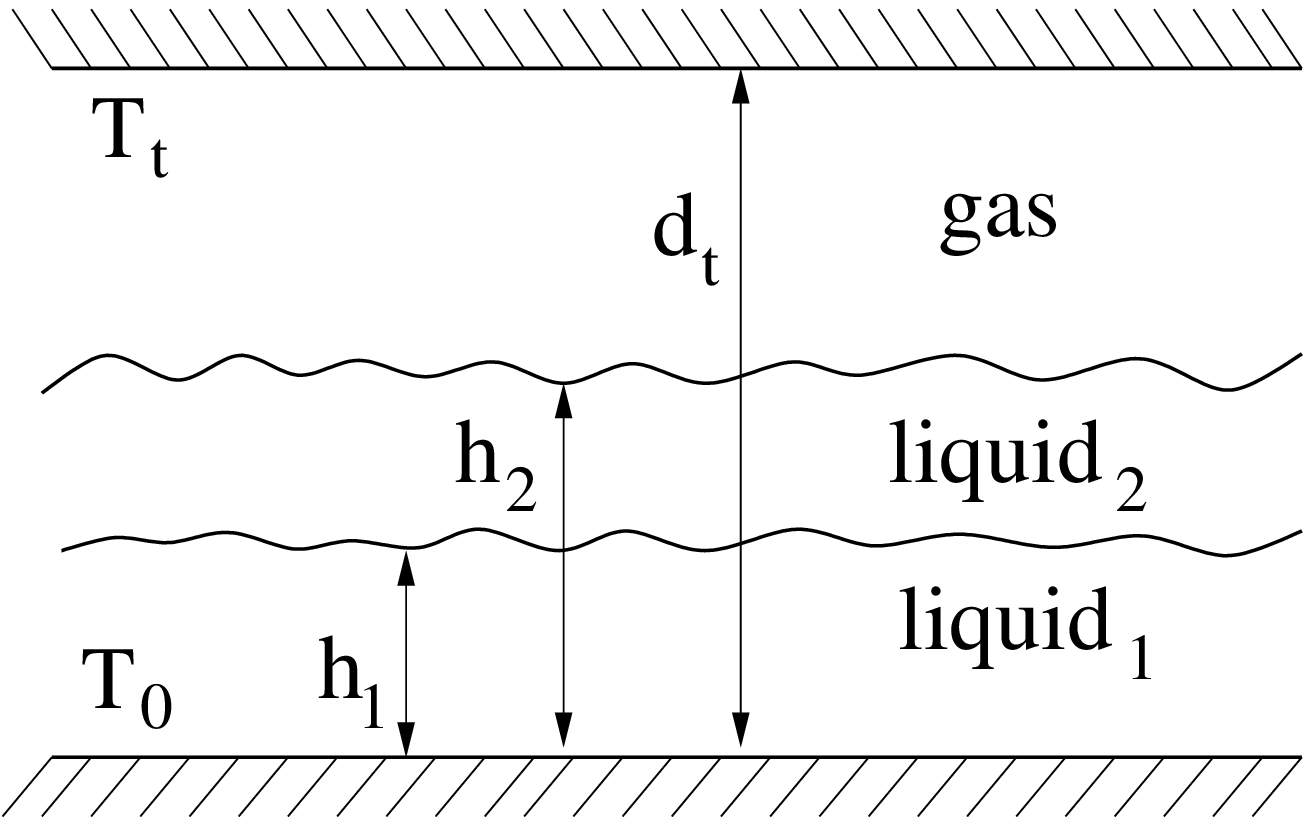}                           
\vspace{4cm}

{\bf\large Fig.\,\ref{fig15}\\
Pototsky et al., J. Chem. Phys.}
\end{center}

\end{document}